\documentclass[preprint,superscriptaddress,amsmath,amssymb,aps,notitlepage,floatfix,longbibliography]{revtex4-1}
\usepackage{graphicx}
\usepackage{color}
\usepackage{comment}

\usepackage{dcolumn}
\usepackage{upgreek}
\usepackage{gensymb}
\usepackage{float}
\usepackage{bm}
\usepackage{xcolor}
\usepackage{physics}
\usepackage{multirow}
\usepackage{array}
\usepackage{upgreek}
\usepackage{ccaption}

\usepackage{xr}
\makeatletter
\newcommand*{\addFileDependency}[1]{
  \typeout{(#1)}
  \@addtofilelist{#1}
  \IfFileExists{#1}{}{\typeout{No file #1.}}
}
\makeatother

\newcommand*{\myexternaldocument}[1]{
    \externaldocument{#1}
    \addFileDependency{#1.tex}
    \addFileDependency{#1.aux}
}

\myexternaldocument{supps_072825}

\begin{document}

\title{Light-programmable reorientation of the crystallographic $c$-axis of Tellurium thin films}


\author{Yuta Kobayashi}
\affiliation{Department of Physics, The University of Tokyo, Tokyo 113-0033, Japan}

\author{Arata Mitsuzuka}
\affiliation{Department of Physics, The University of Tokyo, Tokyo 113-0033, Japan}

\author{Haruo Kondo}
\affiliation{Department of Physics, The University of Tokyo, Tokyo 113-0033, Japan}

\author{Makoto Shoshin}
\affiliation{Department of Physics, The University of Tokyo, Tokyo 113-0033, Japan}

\author{Jun Uzuhashi}
\affiliation{National Institute for Materials Science, Tsukuba 305-0047, Japan}

\author{Tadakatsu Ohkubo}
\affiliation{National Institute for Materials Science, Tsukuba 305-0047, Japan}

\author{Masamitsu Hayashi}
\affiliation{Department of Physics, The University of Tokyo, Tokyo 113-0033, Japan}
\affiliation{Trans-Scale Quantum Science Institute, The University of Tokyo, Bunkyo, Tokyo 113-0033, Japan}

\author{Masashi Kawaguchi}
\affiliation{Department of Physics, The University of Tokyo, Tokyo 113-0033, Japan}

\date{\today}

\begin{abstract}
Tellurium (Te), a two-dimensional material with pronounced structural anisotropy, exhibits exceptional electrical and optical properties that are highly sensitive to its crystallographic orientation. However, conventional synthesis techniques offer limited control over the in-plane alignment of Te's crystallographic $c$-axis, hindering large-scale integration. Here, we report a novel, non-contact method to dynamically manipulate the $c$-axis orientation of Te thin films using linearly polarized picosecond laser pulses. We show that the $c$-axis can be omnidirectionally reoriented perpendicular to the laser polarization, even in initially polycrystalline films. 
This reorientation is fully reversible, allowing for rewritable and spatially selective control of the $c$-axis orientation post-deposition.
Our light-driven approach enables programmable anisotropy in Te, opening new avenues for reconfigurable optoelectronic and photonic devices, such as active metasurfaces and CMOS-compatible architectures.
\end{abstract}
 
\maketitle

\begin{figure}
    \centering
    \includegraphics[width=1.0\linewidth]{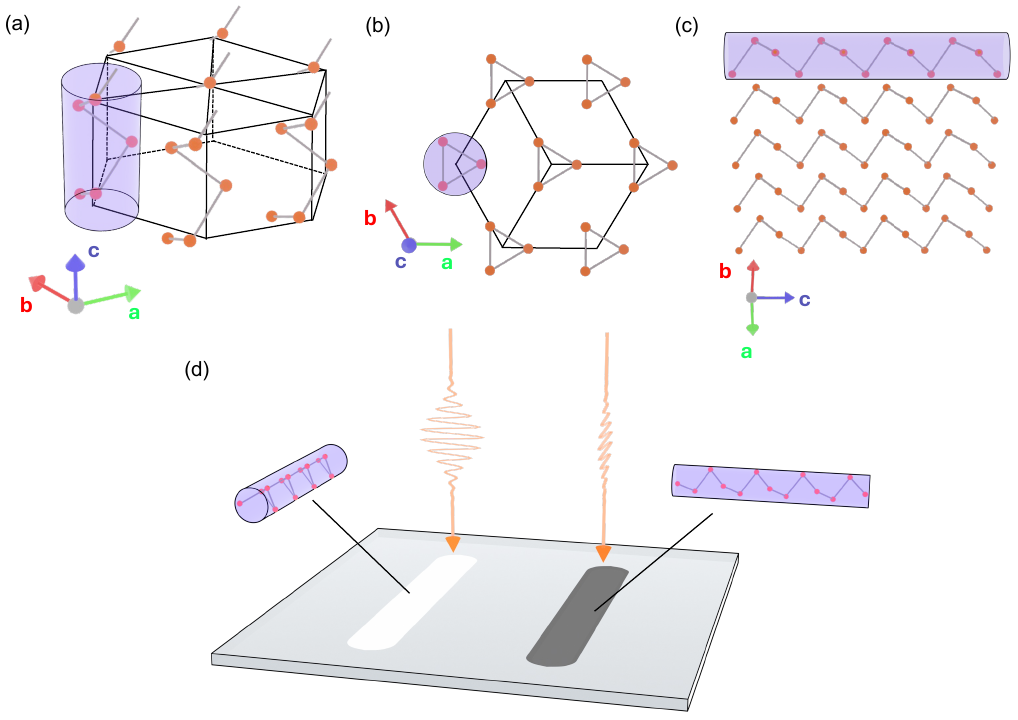}
\caption {\label{fig:Te} 
(a-c) Lattice structure of Te from different viewpoints. The corresponding coordinate axis is shown on the side. The long axis of the cylinder, highlighted with blue, indicates the $c$-axis of Te. (d) Schematic of controlling the direction of the $c$-axis using a linearly polarized laser pulse. The $c$-axis of Te is aligned perpendicular to the laser polarization.}
\end{figure}

\clearpage
Tellurium (Te), a quasi one-dimensional material, exhibits a remarkable combination of physical properties, including structural chirality\cite{dong2020ncomm,sakano2020prl,benmoshe2021science,calavalle2022nmat,ishito2023chirality}, high carrier mobility\cite{wang2018nelec,huang2021jmaterchemc}, Weyl semiconducting characteristics\cite{hirayama2015prl,ideue2019pnas,gatti2020prl,zhang2020pnas,ma2022ncomm}, ferroelectricity\cite{wang2018mathor,zhang2024ncomm} strong mid-infrared absorption\cite{amani2018acsnano,tong2020ncomm} and current induced spin polarization\cite{furukawa2017ncomm,tsirkin2018prb,sahin2018prb,furukawa2021prr}.
These intrinsic features render Te a promising platform for a range of advanced applications\cite{zhu2023frontphys}, such as transistors\cite{wang2018nelec,zhou2018advmater,zhao2020nnano}, radio-frequency diodes\cite{askar2023npj2d} and high-resolution imaging\cite{tong2020ncomm}. The origin of these exceptional properties lies in Te's anisotropic crystal structure: it comprises covalently bonded helical chains weakly coupled via van der Waals forces\cite{vonhippel1948jchemphys,zhu2023frontphys}, as schematically shown in Fig.~\ref{fig:Te}(a-c). This unique structure leads to a strong directional dependence in both carrier transport\cite{rothkirch1969pssb,huang20202dmater,calavalle2022nmat} and optical response\cite{wang2020advmater,guo2022nanoscale}, particularly between the helical axis ($c$-axis) and directions orthogonal to it. 
Conversely, precise control over the orientation of the $c$-axis is of critical importance for optimizing Te-based device functionalities.

While the unique anisotropic properties of Te hold great promise for device applications, the ability to precisely control its crystallographic orientation remains a significant challenge. In many of the studies, Te is synthesized via solution-based methods\cite{wang2018nelec}, which offer a substrate-free route to obtain high-quality Te flakes. However, this approach provides limited control over the in-plane crystal orientation, often necessitating post-selection of suitably aligned flakes for device fabrication and characterization, an obstacle to scalable integration.
Alternative approaches to orienting Te crystals include the use of crystalline substrates\cite{bianco2020nanoscale,zhao2022advmaterinter} or mechanical exfoliation from bulk crystals\cite{churchill2017nanoreslett}. Although these methods offer some control of orientation, the former is inherently constrained by the fixed lattice symmetry of the substrate while the latter is limited by the small sample area attainable. The development of versatile and scalable methods for controlling Te crystal orientation is thus key to advancing its practical applications.

Here, we demonstrate a fundamentally different strategy for manipulating the crystallographic orientation of Te. We show that irradiation of linearly polarized laser pulses can reorient the $c$-axis of Te films to lie perpendicular to the polarization direction, even when the as-deposited films are initially polycrystalline. Remarkably, this laser-induced reorientation is independent of the substrate’s crystallographic orientation and is fully reversible, allowing for programmable control of the $c$-axis orientation upon film deposition.
This light-driven approach, as illustrated in Fig.~\ref{fig:Te}(d), provides dynamic and flexible control over Te crystal alignment, overcoming long-standing limitations associated with substrate constraints and static growth conditions. The ability to reconfigure crystal orientation in a non-contact, spatially resolved manner opens new avenues for integrating Te into advanced optoelectronic and electronic devices, where directional anisotropy plays a pivotal role.

Films made of Te (10, 40 nm)/Al$_2$O$_3$ (1 nm) were deposited on $c$-Al$_2$O$_3$ substrates using molecular beam epitaxy (MBE). 
X-ray diffraction (XRD), scanning electron microscopy (SEM) and cross sectional high-angle annular dark field scanning transmission electron microscopy (HAADF-STEM) were used to characterize the structure of the films. 
The XRD spectra of Te films reveal that the films have a trigonal crystal structure and their $c$-axis is oriented within the film plane.
The HAADF-STEM image of an as-grown Te film is shown in Fig.~\ref{fig:tem}(a). 
To study the crystalline orientation of the film, nanobeam electron diffraction pattern of selected regions in the STEM image are taken.
The electron beam spot size is close to a few atomic radii.
Representative diffraction patterns of the sapphire substrate and the Te film are shown in Figs.~\ref{fig:tem}(c) and \ref{fig:tem}(d), respectively.
The diffraction pattern of the substrate indicates that its $c$-axis points along the surface normal, as illustrated by the orange arrow. 
In contrast, the ring-like pattern (Debye-Scherrer ring) of the Te film suggests that the film is poly-crystalline. 

The Te films were irradiated with a train of linearly polarized picosecond long laser pulses from the film normal. 
The center wavelength and the duration of the laser were 1030 nm and $\sim 4 - 10$ ps, respectively.
The pulse laser was scanned across a region of the Te film. 
The scan rate is 0.05 mm/s (0.1 mm/s) for the 10 nm (40 nm)-thick Te film: the corresponding number of pulses $N$ the film receives is $N = 5.86\times10^6$ ($N = 2.93\times10^6$). In determining $N$, the laser intensity is assumed to be constant inside the laser spot.
We use standard photolithography to pattern the film so that the region irradiated with the laser pulses can be identified.
Figure~\ref{fig:tem}(b) shows a top view of the patterned film.
The bright and dark contrast correspond to the Te film and substrate surface, respectively.
The four regions highlighted with yellow indicate the area where the laser pulses were applied.
The polarization of the laser pulse was set in different directions for the four regions.
Specimen fabricated for TEM observation is indicated by the horizontal lines, parallel to $x$, in Fig.~\ref{fig:tem}(b).

\begin{figure}
	\centering
	\includegraphics[width=1.0\linewidth]{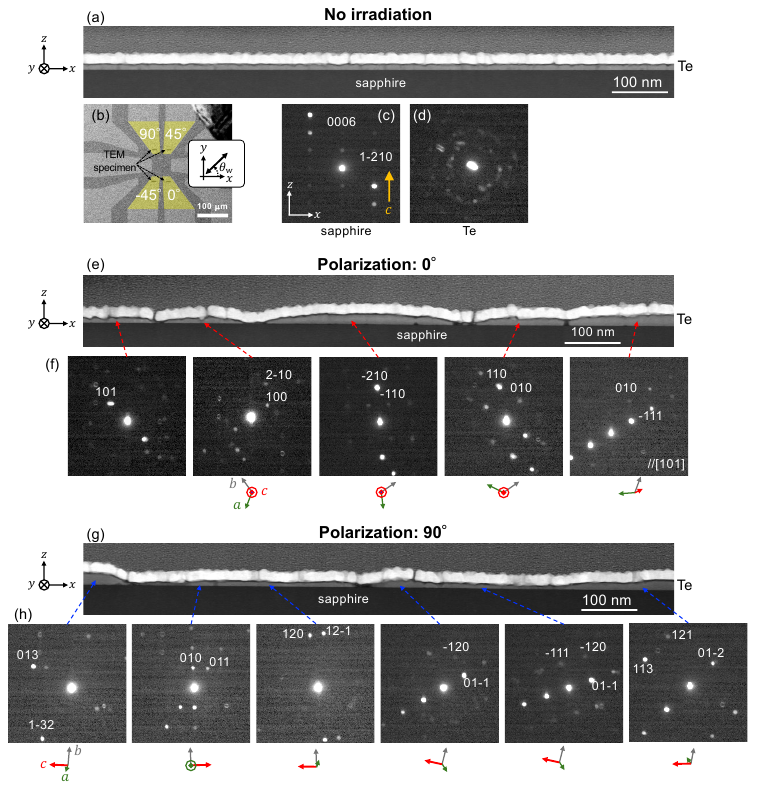}		
	\caption{\label{fig:tem} 
	 (a,c,d) Cross-sectional HAADF-STEM image (a) and the corresponding nanobeam electron diffraction pattern of the sapphire ($c$-Al$_2$O$_3$) substrate (c) and the as-deposited Te film (d). The orange arrow in (c) depicts the orientation of the Al$_2$O$_3$ $c$-axis determined from the diffraction pattern. (b) Optical microscopy image of the patterned film. The region highlighted with yellow represents the area where the film was irradiated with laser pulses. (Caption continues to the next page.)}
\end{figure}
\begin{figure}[t]
  \contcaption{The polarization of the laser pulses with respect to the $x$-axis, i.e. the angle $\theta_\mathrm{w}$, is indicated in each region. The specimen fabricated from the film for TEM observation is indicated by the horizontal lines. (e-h) Cross-sectional HAADF-STEM image (e,g) and the corresponding nanobeam electron diffraction pattern (f,h) of the Te film irradiated with laser pulses with polarization along $x$ ($\theta_\mathrm{w} = 0$ deg) (e,f) and $y$ ($\theta_\mathrm{w} = 90$ deg) (g,h). The arrows below each image in (f) and (h) indicate the crystal axis of Te inferred from the diffraction pattern. The red arrow shows the $c$-axis. For the left most diffraction pattern in (f), we cannot determine the $c$-axis direction of the grain since only the diffraction spots from the (101) orientation are observed. (c,f,h) The Miller indices for the diffraction spots are labeled. (a-h) The Te film thickness is 10 nm. The laser energy used to control the $c$-axis is 20 mJ/cm$^2$.}
\end{figure}

Figures~\ref{fig:tem}(e) and \ref{fig:tem}(g) show the HAADF-STEM images of the Te film after the laser pulse irradiation where the laser polarization is aligned along the $x$-axis ($\theta_\mathrm{w}$ = 0 deg) and $y$-axis ($\theta_\mathrm{w}$ = 90 deg), respectively.
$\theta_\mathrm{w}$ is defined as the angle between the $x$-axis and the polarization of the laser pulses: see Fig.~\ref{fig:tem}(b).
The corresponding nanobeam electron diffraction patterns of regions indicated by the arrows in the STEM images are shown in Figs.~\ref{fig:tem}(f) and \ref{fig:tem}(h).
In the bottom of each image, we show the expected orientation of the crystal axis reconstructed from the diffraction pattern (right handed coordinate axis is assumed).
The red arrow indicates the Te crystal $c$-axis.
We find that the $c$-axis is aligned perpendicular to the laser polarization after the irradiation.
That is, when the laser polarization is set along the $x$-axis [Fig.~\ref{fig:tem}(e,f)], the $c$-axis points along the $y$-axis, and vice versa [Fig.~\ref{fig:tem}(g,h)].
These results clearly show that the polarization of the pulse laser controls the direction of the Te $c$-axis.

\begin{figure}
    \centering
    \includegraphics[width=1.0\linewidth]{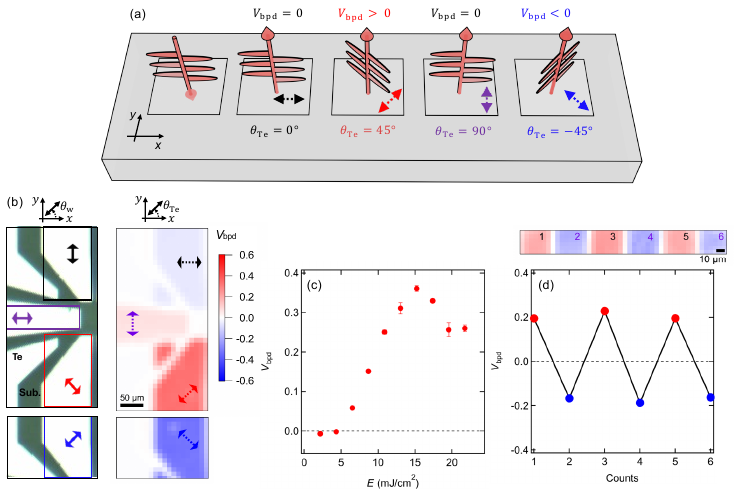}
    \caption{\label{fig:switching} 
    (a) Schematic illustration of the polarization of incident and reflected "probe" light. 
    The polarization of the incident light is illustrated in the left image and that of the reflected light is depicted in the four images shown to the right, in accordance to the birefringence of Te. The $c$-axis of the Te film is indicated by the double-pointed arrow.
    (b) Optical image of the patterned Te film (left panel) and the corresponding two dimensional mapping of $V_\mathrm{bpd}$ (right panel). The double-pointed arrows in the left panel represent the direction of the polarization of the pulse laser whereas the those in the right panel indicate the expected $c$-axis of Te defined by the writing process. The energy of the  pulse laser was 13 mJ/cm$^2$. 
    (c) Pulse laser energy $E$ dependence of $V_\mathrm{bpd}$ with $\theta_{\mathrm{w}}=-45\deg$. The error bars indicate standard deviation of the 9 data points in the two-dimensional mapping.
    (d) Top panel: Two dimensional mapping of $V_\mathrm{bpd}$ for Te film exposed to different numbers of writing. The write count is indicated in each region. The film is irradiated with laser pulses with $\theta_{\textrm{w}}=135 \ (45) \deg$ for odd (even)-numbered writing. The energy of the  pulse laser was 11 mJ/cm$^2$.
    (a-d) All data presented are obtained using a 40 nm-thick Te film.}
\end{figure}

To systematically characterize the pulse laser induced control of the Te $c$-axis, we use optical measurements to determine the direction of the $c$-axis.
Figure~\ref{fig:switching}(a) schematically shows the working principle of the measurements\cite{marui2023prb}.
We first "write" the $c$-axis of the film using a linearly polarized pulse laser.
According to the results shown in Fig.~\ref{fig:tem}, the $c$-axis of the Te film is aligned orthogonal to the polarization of the laser pulses.
For convenience, the angle between the $x$-axis and the expected $c$-axis of the film is labeled as $\theta_\mathrm{Te}$.
($\theta_\mathrm{Te}$ is therefore orthogonal to $\theta_\mathrm{w}$.)
We then use optical reflectometry to study the $c$-axis of the film.
A linearly polarized continuous wave (cw) "probe" laser (wavelength: 633 nm) is incident on the film from the surface normal.
As schematically illustrated in Fig.~\ref{fig:switching}(a), the polarization of incident light is set along the $x$-axis.
The polarization of the reflected light is dependent on the relative orientation of the incident light polarization and the $c$-axis of the Te film due to birefringence.
The birefringent properties of the film\cite{guo2022nanoscale} dictate that the polarization of the reflected light takes the largest (smallest) value when $\theta_\mathrm{Te}$ is $45 +180 n$ deg ($-45 +180 n$ deg) away from the probe light polarization, where $n$ is an integer.
The polarization does not change upon reflection when $\theta_\mathrm{Te}$ is $0 + 90 n$ deg.
Measuring the polarization of the reflected light therefore allows one to determine the direction of the $c$-axis.
The reflected light is guided to a balanced photodetector via a half wave plate and a polarizing beamsplitter.
The signal captured at the balanced photodetector is defined as $V_\mathrm{bpd}$.
$V_\mathrm{bpd}$ is proportional to the difference in the amplitude of the horizontal and vertical components of the light incident on the polarizing beamsplitter.
The difference is normalized by the mean amplitude of the two components.
$V_\mathrm{bpd}$ thus scales with $\theta_\mathrm{Te}$ when $|\theta_\mathrm{Te}| \leq \frac{\pi}{4}$ and provides a measure of the direction of the film's $c$-axis.

The left panels in Fig.~\ref{fig:switching}(b) show optical microscopy images of a patterned Te thin film.
The colored boxes indicate the areas that were irradiated with "write" laser pulses with different polarization (i.e. with different values of $\theta_\mathrm{w}$).
The double sided solid arrows indicate the direction of the polarization of the write pulses.
The right panels in Fig.~\ref{fig:switching}(b) show a corresponding two-dimensional mapping of $V_\mathrm{bpd}$.
As is evident, $V_\mathrm{bpd}$ takes a minimum and maximum when $\theta_\mathrm{w}$ is 45 deg and -45 deg, respectively.
For $\theta_\mathrm{w} = 0$ and 90 deg, $V_\mathrm{bpd}$ is near zero.
In accordance with the birefringent properties of the film [Fig.~\ref{fig:switching}(a)], these results indicate that the $c$-axis of the Te film, indicated by the double sided arrows in the right panels of Fig.~\ref{fig:switching}(b), is orthogonal to the polarization of the write pulses.
These results are consistent with the analyses of the HAADF-STEM images: see Fig.~\ref{fig:tem}. 
We further verified that $V_\mathrm{bpd}$ depends not on the write polarization angle $\theta_\mathrm{w}$ itself, but rather on the relative angle between the resulting crystal axis and the polarization of the probe light.
Such angular dependence of $V_\mathrm{bpd}$ is found for all regions with different $\theta_\mathrm{Te}$ and is well reproduced by a model that assumes uniaxial optical anisotropy, validating our interpretation.
These findings confirm that the writing process is omnidirectional in nature, that is, it is independent of the substrate orientation and governed solely by the polarization of the write pulse. 
Note that the results presented in Figs.~\ref{fig:tem} and \ref{fig:switching} are obtained using Te films with different thicknesses, suggesting that the writing of the $c$-axis is not significantly influenced by the film thickness as well.

The pulse laser energy $E$ dependence of $V_\mathrm{bpd}$ is shown in Fig.~\ref{fig:switching}(c).
$V_\mathrm{bpd}$ exhibits a threshold at $E \sim 5$ mJ/cm$^2$ and takes a maximum at $E \sim  15$ mJ/cm$^2$, beyond which it tends to decrease with increasing power.
Such non-monotonic trend suggests that the degree of $c$-axis alignment is optimized at a specific excitation energy. 
We infer that the alignment process can be caused by a laser-induced selective melting-recrystallization process, governed by the strong linear dichroism of Te.
The anisotropic electronic structure and the pronounced birefringence/dichroism of Te\cite{guo2022nanoscale} lead to stronger absorption when the light polarization is aligned with the $c$-axis. 
Note that the HAADF-STEM images (Fig.~\ref{fig:tem}(a,d)) and X-ray diffraction spectrum 
of the as-grown film show that the film is polycrystalline. 
The orientation of the $c$-axis of the film, defined by the orientation of the nano-crystals (or grains), is thus initially randomly distributed.
At intermediate pulse energies  $5 \leq E \leq 15$ mJ/cm$^2$, we consider only the nanocrystals with their $c$-axis parallel to the polarization of the pulse laser absorb sufficient energy to melt. 
Upon cooling, these regions recrystallize. 
Crucially, however, the surrounding unmelted nanocrystals with perpendicular $c$-axis orientation may serve as structural templates, guiding the re-nucleation and oriented regrowth in preferred orientations orthogonal to the laser polarization. 
Repeating this process over multiple pulses progressively enhances the fraction of domains aligned perpendicular to the laser polarization. 
At higher pulse energies ($E \geq 15$ mJ/cm$^2$), thermal excitation becomes indiscriminate, melting crystals regardless of their initial orientation, thereby diminishing the directionality of the reorientation process.

In addition to thermally driven dynamics, we speculate that the intense electric field of the picosecond laser pulse may transiently polarize the molten or near-molten Te regions. In analogy with all-optical magnetization switching in magnetic media\cite{stanciu2007prl1,lambert2014science,gorchon2016prb}, the intense electric field of the picosecond laser pulse may act synergistically with the anisotropic optical response and local crystallographic order to energetically favor certain orientations during resolidification. The possibility that residual crystalline regions or the laser-induced field environment collectively function as a self-assembled orientation field represents an intriguing direction for future investigation\cite{zuo2019acsapplmaterinter,kim2023nanolett,kadoguchi2023nanolett}.


Finally, we show that the writing of the $c$-axis with laser pulses can be conducted repetitively, that is, the $c$-axis can be overwritten by laser pulses with different polarization.
The top panel of Fig.~\ref{fig:switching}(d) shows a two-dimensional mapping of $V_{\textrm{bpd}}$ of a Te film where different parts of the film were exposed with different numbers of writing.
The color scale is the same as that of Fig.~\ref{fig:switching}(b).
Here we set the pulse laser energy to 11 mJ/cm$^2$.
The areas labeled 1 to 6 in Fig.~\ref{fig:switching}(d), top panel, were first exposed to write laser pulses with polarization along $\theta_\textrm{w} = -45$ deg.
Regions 2 to 6 were then irradiated with laser pulses with polarization along $\theta_\textrm{w}=45$ deg, followed by exposure of laser pulses with $\theta_\textrm{w} = -45$ deg to regions 3 to 6, and so on.
With this process, region $i$ ($i = 1 \sim 6$) is exposed to the writing process $i$ times, in which each successive writing alters its polarization between $\theta_\textrm{w} = -45$ deg and $\theta_\textrm{w}=45$ deg.
The color contrast of the top panel of Fig.~\ref{fig:switching}(d) shows that $V_{\textrm{bpd}}$ is positive for odd number of writing and is negative for even number of writing.
The absolute value of $V_{\textrm{bpd}}$ is nearly the same ($\sim 0.2$) after each writing process, consistent with the results shown in Fig.~\ref{fig:switching}(c) and the pulse laser energy used.
These results thus demonstrate that successive writing with different laser polarization can switch the Te $c$-axis repeatedly.

In summary, we have demonstrated that the crystallographic $c$-axis of Te thin films can be reoriented using linearly polarized picosecond laser pulses. The $c$-axis aligns perpendicular to the polarization direction, even in polycrystalline films, and can be arbitrarily rewritten upon further illumination. 
This light-driven, substrate orientation-independent approach enables spatially resolved control of crystallographic anisotropy.
Such a mechanism not only complements substrate-guided growth techniques, but also introduces an approach to engineering anisotropy in low-dimensional materials with pronounced in-plane optical and electronic anisotropy,
e.g. black phosphorus\cite{xia2014ncomm}.
These capabilities offer a scalable route toward reconfigurable optoelectronic architectures, including active metasurfaces\cite{yu2011science,shaltout2019science,dorrah2022science}, polarized infrared imaging\cite{tong2020ncomm}, and CMOS-compatible platforms\cite{wang2018nelec}.



\section*{Acknowledgments}
This work was partly supported by JST CREST (JPMJCR19T3), JSPS KAKENHI (Grant Numbers 21K13863, 23H00176), MEXT Initiative to Establish Next-generation Novel Integrated Circuits Centers (X-NICS) and Cooperative Research Project Program of RIEC, Tohoku University. A.M. and M.S. were supported by Materials Education program for the future leaders in Research, Industry, and Technology (MERIT), The University of Tokyo.

\section*{Conflict of Interest}
The authors declare no competing interests.

\section*{Author contributions}
M.K. and M.S. found the effect. A.M. and H.K. deposited the films and fabricated the samples. Y.K., A.M., H.K. performed the optical measurements with help from M.S. and M.K. J.U. and T.O. took the HAADF-STEM and SEM images. Y.K. and M.H. wrote the manuscript with substantial inputs from all authors.

\section*{Data Availability Statement}
The data that support the findings of this study are available from the corresponding author upon reasonable request.

\section*{Keywords}
tellurium, anisotropic materials, birefringence, dichroism, ultrashort laser pulses

\clearpage
\bibliography{refs_061325}

\end{document}